\newcommand{\minpoint}{\mbox{$'\mskip-4.7mu.\mskip0.8mu$}}
\documentstyle[emulateapj,epsf]{article}
\submitted{Accepted for publication in the Astrophysical Journal, Letters}

\begin{document}

% \submitted{} 

\title{THE DISCOVERY OF A VERY NARROW-LINE STAR FORMING OBJECT
       AT A REDSHIFT OF 5.66\altaffilmark{1,2}}

\author{Yoshiaki Taniguchi  \altaffilmark{3},
        Masaru Ajiki        \altaffilmark{3},
        Takashi Murayama    \altaffilmark{3},
        Tohru Nagao         \altaffilmark{3},
        Sylvain Veilleux    \altaffilmark{4},
        David B. Sanders    \altaffilmark{5, 6},
        Yutaka Komiyama     \altaffilmark{7},
        Yasuhiro Shioya     \altaffilmark{3},
        Shinobu S. Fujita   \altaffilmark{3},
        Yuko Kakazu         \altaffilmark{5},
        Sadanori Okamura    \altaffilmark{8,9},
	Hiroyasu Ando       \altaffilmark{7},
        Tetsuo Nishimura    \altaffilmark{7},
        Masahiko Hayashi    \altaffilmark{7},
        Ryusuke Ogasawara   \altaffilmark{7}, \&
        Shin-ichi Ichikawa  \altaffilmark{10}
	}

\altaffiltext{1}{Based on data collected at 
	Subaru Telescope, which is operated by 
	the National Astronomical Observatory of Japan.}
\altaffiltext{2}{Based on data collected at
        the W. M. Keck Observatory, which is operated as a scientific 
	partnership among the California Institute of Technology, the 
	University of California, and the National Aeronautics and Space 
	Administration.}
\altaffiltext{3}{Astronomical Institute, Graduate School of Science,
        Tohoku University, Aramaki, Aoba, Sendai 980-8578, Japan}
\altaffiltext{4}{Department of Astronomy, University of Maryland, 
        College Park, MD 20742}
\altaffiltext{5}{Institute for Astronomy, University of Hawaii,
        2680 Woodlawn Drive, Honolulu, HI 96822}
\altaffiltext{6}{Max-Planck Institut fur Extraterrestriche Physik, D-85740, 
        Garching, Germany}
\altaffiltext{7}{Subaru Telescope, National Astronomical Observatory,
        650 N. A'ohoku Place, Hilo, HI 96720}
\altaffiltext{8}{Department of Astronomy, Graduate School of Science,
        University of Tokyo, Tokyo 113-0033, Japan}
\altaffiltext{9}{Research Center for the Early Universe, School of Science,
        University of Tokyo, Tokyo 113-0033, Japan}
\altaffiltext{10}{National Astronomical Observatory, 2-21-1 Osawa,
	Mitaka, Tokyo 181-8588, Japan}

\begin{abstract} 
We report on the discovery of a very narrow-line star forming object
beyond redshift of 5. 
Using the prime-focus camera, Suprime-Cam, on the 8.2 m Subaru
telescope together with a narrow-passband filter centered at
$\lambda_{\rm c}$ = 8150 \AA\ with passband of $\Delta\lambda$ = 120
\AA, we have obtained a very deep image of the field surrounding the
quasar SDSSp J104433.04$-$012502.2 at a redshift of 5.74.
Comparing this image with optical broad-band images, we have found an
object with a very strong emission line. Our follow-up optical
spectroscopy has revealed that this source is at a redshift of
$z=5.655\pm0.002$, forming stars at a rate $\sim 13 ~ h_{0.7}^{-2} ~ M_\odot$
yr$^{-1}$.
Remarkably, the velocity dispersion of Ly$\alpha$-emitting
gas is only 22 km s$^{-1}$.  Since a blue half of the Ly$\alpha$ emission 
could be absorbed by neutral hydrogen gas, perhaps in the system,
a modest estimate of the velocity dispersion may be $\gtrsim$ 44 km s$^{-1}$.
Together with a linear size of 7.7 $h_{0.7}^{-1}$ kpc, 
we estimate a lower limit of
the dynamical mass of this object to be $\sim 2 \times
10^9 M_\odot$. 
It is thus suggested that LAE J1044$-$0123 is a star-forming 
dwarf galaxy (i.e., a subgalactic object or a building block)
beyond redshift 5 although 
we cannot exclude a possibility that most Ly$\alpha$ emission is
absorbed by the red damping wing of neutral intergalactic matter.
\end{abstract}


\section{INTRODUCTION}

Searches for Ly$\alpha$ emitters (hereafter LAEs) at high redshift are very
useful in investigating the early star formation history of galaxies
(Partridge \& Peebles 1967). Recent progress in the observational 
capability of 10-m class optical telescopes has enabled us to
discover more than a dozen of Ly$\alpha$ emitting galaxies beyond
redshift 5 (Dey et al. 1998; Weymann et al. 1998; Hu et al. 1999, 
2001, 2002; Dawson et al. 2001, 2002; Ellis et al. 2001; Ajiki et al.
2002; see also Spinrad et al. 1998; Rhoads \& Malhotra 2001).
The most distant object known to date is HCM-6A at $z=6.56$ (Hu et al. 2002).
These high-$z$ Ly$\alpha$ emitters can be also utilized to investigate
physical properties of the intergalactic medium (IGM) because 
the epoch of cosmic reionization ($z_{\rm r}$) is considered to be close to
the redshifts of high-$z$ Ly$\alpha$ emitters; i.e., $z_{\rm r} \sim
6$ -- 7 (Djorgovski et al. 2001; Becker et al. 2001; Fan et al. 2002).
In other words, emission-line fluxes of the Ly$\alpha$ emission
from such high-$z$ galaxies could be absorbed by neutral hydrogen
in the IGM if neutral gas clouds are located between the source and us
(Gunn \& Peterson 1965; Miralda-Escud\'e 1998; Miralda-Escud\'e \& Rees 1998;
Haiman 2002 and references therein).
Therefore, careful investigations of Ly$\alpha$ emission-line properties of
galaxies with $z > 5$ provides very important clues simultaneously
both on the early star formation history of galaxies and on the physical
status of IGM at high redshift.

The first step is to look for a large number of Ly$\alpha$ emitter 
candidates at high redshift through direct imaging surveys.
The detectability of high-redshift objects is significantly increased
if they are hosts to recent bursts of star formation which ionize the
surrounding gas and result in strong emission lines like the hydrogen
recombination line Ly$\alpha$.
These emission-line objects can be found in principle through
deep optical imaging with narrow-passband filters customized to the
appropriate redshift.  Indeed, recent attempts with the Keck 10 m
telescope have revealed the presence of Ly$\alpha$ emitters in blank
fields at high redshift (e.g., Cowie \& Hu 1998; Hu et al. 2002).
These recent successes have shown
the great potential of narrow-band imaging surveys with 8-10 m
telescopes in the search for high-$z$ Ly$\alpha$ emitters.
It is also worthwhile noting that subgalactic populations at high redshift
have been recently found thanks to the gravitational lensing
(Ellis et al. 2001; Hu et al. 2002).

In an attempt to find star-forming objects at $z \approx 5.7$,
we have carried out a very deep optical imaging survey in the field
surrounding the quasar SDSSp J104433.04$-$012502.2 at redshift of
5.74\footnote{
The discovery redshift was $z=5.8$ (Fan et al. 2000).
Since, however, the subsequent optical spectroscopic observations
suggested a bit lower redshift; $z=5.73$ (Djorgovski et al. 2001)
and $z=5.745$ (Goodrich et al. 2001), we adopt $z=5.74$ in this Letter.}
(Fan et al. 2000; Djorgovski et al. 2001; Goodrich et al. 2001),
using  Suprime-Cam (Miyazaki et al. 1998), the wide-field
($34^\prime \times 27^\prime$ with a 0.2 arcsec/pixel resolution)
prime-focus camera on the 8.2 m Subaru telescope (Kaifu 1998).
In this Letter, we report on our discovery of a very narrow-line 
star-forming system at $z \approx 5.7$.

\section{OBSERVATIONS}

\subsection{Optical Imaging}

In this survey, we used the narrow-passband filter, NB816, centered on 
8150 \AA\ with a passband of $\Delta\lambda$(FWHM) = 120 \AA;
the central wavelength corresponds to a redshift of 5.70 for
Ly$\alpha$ emission.  We also used broad-passband filters, $B$, $R_{\rm C}$,
$I_{\rm C}$, and $z^\prime$. A summary of the imaging observations is
given in Table 1. All of the observations were done under photometric
condition and the seeing size was between 0.7 arcsec and 1.3 arcsec
during the run.
Note that we analyzed only two CCD chips, in which quasar SDSSp
J104433.04$-$012502.2 is present, to  avoid delays for follow-up spectroscopy.
The CCD data were reduced and combined using $IRAF$ and the mosaic-CCD data
reduction software developed by Yagi et al. (2002). 
Photometric and spectrophotometric
standard stars used in the flux calibration are SA101 for the
$B$, $R_{\rm C}$, and $I_{\rm C}$ data, and GD 108,
GD 58 (Oke 1990), and PG 1034+001 (Massey et al. 1988)
for the NB816 data. The $z^\prime$ data were
calibrated by using the magnitude of SDSSp J104433.04$-$012502.2
(Fan et al. 2000).

The total size of the field is 11\minpoint67 by 11\minpoint67, corresponding
to a solid angle of $\approx$ 136 arcmin$^{2}$. The volume probed by the
NB816 imaging has (co--moving) transverse dimensions of 27.56
$h_{0.7}^{-1}\times 27.56 h_{0.7}^{-1}$ Mpc$^2$, and the half--power
points of the filter correspond to a co--moving depth along the line
of sight of 44.34 $h_{0.7}^{-1}$ Mpc ($z_{\rm min} \approx 5.653$ and
$z_{\rm max} \approx 5.752$; note that the transmission curve of our
NB816 filter has a Gaussian-like shape). Therefore, a total volume of
$3.4 \times 10^4 h_{0.7}^{-3}$ Mpc$^{3}$ is probed in our NB816 image.
Here, we adopt a flat universe with $\Omega_{\rm matter} = 0.3$,
$\Omega_{\Lambda} = 0.7$, and $h=0.7$ where $h = H_0/($100 km s$^{-1}$
Mpc$^{-1}$).

Source detection and photometry were performed using SExtractor
version 2.2.1 (Bertin, \& Arnouts 1996).  
Our detection limit (a 3$\sigma$ detection
within a $2^{\prime\prime}$.8 diameter aperture) for each band is
listed in Table 1. As for the source detection in the NB816
image, we used a criterion that a source must be a 13-pixel connection
above 5$\sigma$ noise level.  Adopting the criterion for the NB816
excess, $I_{\rm C} - NB816 > 1.0$ mag, we have found two strong
emission-line sources. Our follow-up optical spectroscopy of these
sources reveals that one source found
at $\alpha$(J2000)=10$^{\rm h}$
44$^{\rm m}$ 27$^{\rm s}$ and $\delta$(J2000)=$-01^\circ$ 23$^\prime$
45$^{\prime\prime}$ (hereafter LAE J1044$-$0123)
is a good candidate
to be a subgalactic object at high redshift\footnote{Another source
has been identified as a Ly$\alpha$ emitter at $z=5.687$ (Ajiki et al.
2002)}. Its AB magnitude in the $NB816$ band is 24.73.
The optical thumb-nail images of
LAE J1044$-$0123 are given in Fig. 1.  As shown in this figure, LAE
J1044$-$0123 is seen clearly only in the NB816 image. Although it is
seen in the $I_{\rm C}$ image, its flux is below the $3\sigma$ noise
level.  The observed equivalent width is $EW_{\rm obs} > 238$ \AA.
The NB816 image reveals that LAE J1044$-$0123 is spatially extended;
its angular diameter is 1.6 arcsec (above the 2$\sigma$ noise
level). The size of the point spread function in the NB816 image is
0.90 arcsec. Correcting for this spread, we obtain an angular
diameter of 1.3 arcsec.  

\subsection{Optical Spectroscopy}

Our optical spectroscopy was made by using the Keck II Echelle 
Spectrograph and Imager (ESI: Sheinis et al. 2000) on 2002 March 15 (UT).
We used the Echelle mode with the
slit width of 1 arcsec, resulting in a spectral resolution $R \simeq
3400$ at 8000 \AA. The integration time was 1800 seconds.  The
spectrum of LAE J1044$-$0123, shown in Fig. 2, presents a narrow
emission line at $\lambda = 8090$ \AA. This is the only emission line
that was detected within the ESI wavelength range (from 4000 \AA\ to
9500 \AA).  This line may be either Ly$\alpha$ or [O {\sc
ii}]$\lambda$3727. The emission-line profile appears to show a
sharper cutoff at wavelengths shortward of the line peak, providing
some evidence that this line is Ly$\alpha$.  A stronger argument in
favor of this line identification comes from the lack of structure in
the profile. If this line were [O {\sc ii}] emission, the redshift
would be $z \approx 1.17$. Since the [O {\sc ii]} feature is a doublet
line of [O {\sc ii}]$\lambda$3726.0 and [O {\sc ii}]$\lambda$3728.8,
the line separation would be larger than 6.1 \AA\ and the lines
would be resolved in the ESI observations.
Further, if the line were H$\beta$, [O {\sc iii}]$\lambda$4959,
[O {\sc iii}]$\lambda$5007, or H$\alpha$ line, we would detect
some other emission lines in our spectrum.
Therefore, we conclude that the emission line at 8090 \AA\ is
Ly$\alpha$, giving a redshift of 5.655$\pm$0.002.

\section{RESULTS AND DISCUSSION}

\subsection{Star Formation Activity in LAE J1044$-$0123}

The rest-frame equivalent width of Ly$\alpha$ emission
becomes $EW_0 > 36$ \AA.
Our Keck/ESI spectrum gives the observed Ly$\alpha$ flux of $f$(Ly$\alpha$) =
$(1.3 \pm 0.1) \times 10^{-17}$ ergs cm$^{-2}$ s$^{-1}$
and the rest-frame equivalent width of Ly$\alpha$ emission
$EW_0 > 36$ \AA.
On the other hand, our NB816 magnitude of LAE J1044$-$0123
gives $f$(Ly$\alpha$) $\simeq 4.1 \times 10^{-17}$ ergs cm$^{-2}$ s$^{-1}$,
being higher than by a factor of 3 than the Keck/ESI flux.
Since the Keck/ESI spectrum was calibrated by a single measurement
of a spectroscopic standard star, HZ 44, the photometric accuracy
may not be good. Further, our slit width may not cover the entire 
Ly$\alpha$ nebula of LAE J1044$-$0123. 
Therefore, we use the NB816-based flux to estimate the star formation rate.
The NB816 flux gives the Ly$\alpha$ luminosity $L$(Ly$\alpha$) $\simeq 
1.4  \times 10^{43} ~ h_{0.7}^{-2}$ ergs s$^{-1}$.
Using the relation $SFR = 9.1 \times 10^{-43}
L({\rm Ly}\alpha) ~ M_\odot {\rm yr}^{-1}$ (Kennicutt 1998;
Brocklehurst 1971), we obtain
$\sim 13 ~h_{0.7}^{-2} ~ M_\odot$ yr$^{-1}$.
This is a lower limit because no correction was made for possible
internal extinction by dust grains in the system. The lack of UV
continuum from this object prevents us from determining the importance
of this effect.

The most intriguing property of LAE J1044$-$0123 is that the observed
emission-line width (full width at half maximum; FWHM) of redshifted
Ly$\alpha$ is only 2.2 $\pm$ 0.3 \AA. Since the instrumental spectral
resolution is 1.7 $\pm$ 0.1 \AA, the intrinsic width is only 1.4 $\pm$
0.5 \AA; note that this gives a upper limit because the line is barely
resolved. It corresponds to $FWHM_{\rm obs} \simeq$ 52 $\pm$ 19 km s$^{-1}$ or a
velocity dispersion $\sigma_{\rm obs} = FWHM_{\rm obs}/(2 \sqrt{2 {\rm ln} 2}) \simeq$
22 km s$^{-1}$. This value is comparable to those of luminous globular
clusters (Djorgovski 1995). 

It is interesting to compare the observational properties of
LAE J1044$-$0123 with similar LAEs at $z \gtrsim 5$.
For this comparison, we choose Abell 2218 a (Ellis et al. 2001),
LAE J1044$-$0130 (Ajiki et al. 2002), and  J123649.2+621539 (Dawson et al. 2002)
because these objects were also observed using KecK/ESI.
A summary is given in Table 2. This comparison shows that
LAE J1044$-$0123 has the narrowest line width that may be
roughly comparable to that of Abell 2218 a although 
the mass of Abell 2218, $\sim 10^6 M_\odot$, is much smaller
than that of LAE J1044$-$0123 (see next subsection).
Another important point appears that the line profile of 
LAE J1044$-$0123 does not show intense red wing emission which
is evidently seen in those of the other three LAEs.
The diversity of the observational properties of these LAEs
suggest that the H {\sc i} absorptions affect significantly
the visibility of the Ly$\alpha$ emission line.
Further, the contribution of superwinds may be different from
LAE to LAE.

\subsection{What is LAE J1044$-$0123 ?}

Now a question arises as; ^^ ^^ What is LAE J1044$-$0123 ?".
There are two alternative ideas: (1) LAE J1044$-$0123 is a part of 
a giant system and we observe only the bright star-forming clump, or
(2) LAE J1044$-$0123 is a single star-forming system.
Solely from our observations, we cannot judge which is the case.
If this is the first case, LAE J1044$-$0123 may be similar to 
Abell 2218 a found by Ellis et al. (2001). One problem in this 
interpretation seems that the spatial extension, $\sim$ 7.7 kpc,
of LAE J1044$-$0123 is fairly large for such a less-massive system.
Therefore, adopting the second case, it seems important to 
investigate possible dynamical status of LAE J1044$-$0123
for future consideration.

If a source is surrounded by neutral hydrogen, Ly$\alpha$ photons emitted
from the source are heavily scattered. Furthermore, the red damping wing of
the Gunn-Peterson trough could also suppress the Ly$\alpha$ emission line
(Gunn \& Peterson 1965; Miralda-Escud\'e
1998; Miralda-Escud\'e \& Rees 1998; Haiman 2002 and references therein).
If this is the case for LAE J1044$-$0123, we may see only a part of 
the Ly$\alpha$ emission. Haiman (2002) estimated that only 8\% of the
Ly$\alpha$ emission is detected in the case of HCM-6A at $z=6.56$
found by Hu et al. (2002).
However, the observed Ly$\alpha$ emission-line profile of LAE J1044$-$0123
shows the sharp cutoff at wavelengths shortward of the line peak.
This property suggests that the H {\sc i} absorption is dominated by 
H {\sc i} gas in the system rather than that in the IGM 
Therefore, it seems reasonable to adopt that blue half of the
Ly$\alpha$ emission could be absorbed in the case of LAE J1044$-$0123. 
Then we estimate a modest estimate of the velocity dispersion,
$\sigma_0 \sim 2 \sigma_{\rm obs} \sim$ 44 km s$^{-1}$. 
Given the diameter of this object probed by the
Ly$\alpha$ emission,  $D \simeq 7.7 h_{0.7}^{-1}$ kpc, 
we obtain the dynamical timescale of $\tau_{\rm dyn} \sim
D/\sigma_0 \sim 1.7 \times 10^8$ yr. This would give a upper limit of
the star formation timescale in the system; i.e., $\tau_{\rm SF}
\lesssim \tau_{\rm dyn}$. However, if the observed diameter is
determined by the so-called Str\"omgren sphere photoionized by 
a central star cluster, it is not necessary to adopt $\tau_{\rm SF}
\sim \tau_{\rm dyn}$. It seems more appropriate to adopt a shorter
timescale for such a high-$z$ star-forming galaxies, e.g., 
$\tau_{\rm SF} \sim 10^7$ yr, as adopted for HCM-6A at $z \approx 6.56$
(Hu et al. 2002) by Haiman (2002).
One may also derive a dynamical mass $M_{\rm dyn} =
(D/2) \sigma_0 ^2 G^{-1} \sim 2 \times 10^9 M_\odot$ (neglecting possible
inclination effects). 

At the source redshift, $z=5.655$, the mass of a dark matter halo which 
could collapse is estimated as
$M_{\rm vir} \sim 9 \times 10^6 r_{\rm vir, 1}^3
h_{0.7}^{-1} ~ M_\odot$
where $r_{\rm vir, 1}$ is the Virial radius in units of 1 kpc
[see equation (24) in Barkana 
\& Loeb (2001)]. If we adopt $r_{\rm vir} = D/2$ = 3.85 kpc,
we would obtain $M_{\rm vir} \sim 5 \times 10^8 ~ M_\odot$.
However, the radius of dark matter halo could be ten times as long as
$D/2$. If this is the case, we obtain $M_{\rm vir} \sim 5 \times 10^{10}
~ M_\odot$ and $\sigma_0 \sim 75$ km s$^{-1}$.
Comparing this velocity dispersion with the observed one, we estimate
that the majority of Ly$\alpha$ emission would be absorbed by neutral
hydrogen. 

The most important issue related to LAE J1044$-$0123 seems 
how massive this source is; i.e., $\sim 10^9 M_\odot$ or more
massive than $10^{10} M_\odot$.
If the star formation timescale is as long as the dynamical one,
the stellar mass assembled in LAE J1044$-$0123 at $z=5.655$ exceeds 
$10^9 ~ M_\odot$,  being comparable to the
nominal dynamical mass, $M_{\rm dyn} \sim 2 \times 10^9 ~ M_\odot$.
Since it is quite unlikely that most mass is assembled to form stars
in the system, the dark matter halo around LAE J1044$-$0123 would be 
more massive by one order of magnitude at least than the above stellar
mass. If this is the case, we could miss the majority of the Ly$\alpha$ 
emission and the absorption cloud be attributed to the red damping
wing of neutral hydrogen in the IGM.
Since the redshift of LAE J1044$-$0123 ($z=5.655$) is close to that
of SDSSp J104433.04$-$012502.2 ($z=5.74$), it is possible that 
these two objects are located at nearly the same cosmological distance.
The angular separation between LAE J1044$-$0123 and SDSSp 
J104433.04$-$012502.2, 113 arcsec, corresponds to the linear
separation of 4.45 $h_{0.7}^{-1}$ Mpc. 
The Str\"omgren radius of SDSSp J104433.04$-$012502.2 can be estimated
to be $r_{\rm S} \sim 6.3 (t_{\rm Q}/2\times 10^7 ~ {\rm yr})^{1/3}$
Mpc using equation (1)
in Haiman \& Cen (2002) where $t_{\rm Q}$ is the lifetime of the quasar
(see also Cen \& Haiman 2000).
Even if this quasar is amplified by a factor of 2 by the gravitational
lensing (Shioya et al. 2002), we obtain $r_{\rm S} \sim 4.9$ Mpc. 
Therefore, it seems likely that the IGM around LAE J1044$-$0123
may be ionized completely. If this is the case, we cannot expect that
the Ly$\alpha$ emission of LAE J1044$-$0123 is severely absorbed
by the red damping wing emission. 
In order to examine which is the case, $L$-band spectroscopy is strongly
recommended because the redshifted [O {\sc iii}]$\lambda$5007 emission
will be detected at 3.33 $\mu$m. However, we need
the James Webb Space Telescope
to complete it.

%-----------------------------------------------------
%    Table 1
%-----------------------------------------------------
\begin{deluxetable}{lcccc}
\tablenum{1}
%\tabletypesize{\scriptsize}
\tablecaption{A journal of imaging observations}
\tablewidth{0pt}
\tablehead{
\colhead{Band} & 
\colhead{Obs. Date (UT)} & 
\colhead{Total Integ. Time (s)}  &
\colhead{$m_{\rm lim}$(AB)\tablenotemark{a}} &
\colhead{FWHM$_{\rm star}$\tablenotemark{b} (arcsec)} 
}
\startdata
$B$         & 2002 February 17      &  1680 & 26.6 & 1.2 \\
$R_{\rm C}$ & 2002 February 15, 16  &  4800 & 26.2 & 1.4 \\
$I_{\rm C}$ & 2002 February 15, 16  &  3360 & 25.9 & 1.2 \\
$NB816$     & 2002 February 15 - 17 & 36000 & 26.0 & 0.9 \\
$z'$        & 2002 February 15, 16  &  5160 & 25.3 & 1.2 \\
\enddata
\tablenotetext{a}{The limiting magnitude (3$\sigma$) with a
2.8$^{\prime\prime} \phi$ aperture. }
\tablenotetext{b}{The full width at half maximum of stellar
objects in the final image}
\end{deluxetable}

%-----------------------------------------------------
%    Table 2
%-----------------------------------------------------
\begin{deluxetable}{lcccc}
\tablenum{2}
%\tabletypesize{\scriptsize}
\tablecaption{Comparison with similar LAEs beyond redshift 5}
\tablewidth{0pt}
\tablehead{
\colhead{Name} &
\colhead{Redshift}  &
\colhead{$f$(Ly$\alpha$)\tablenotemark{a}}  &
\colhead{$FWHM$(Ly$\alpha$)\tablenotemark{b}}  &
\colhead{Ref.\tablenotemark{c}}
}
\startdata
LAE J1044$-$0123 & 5.655 & 4.1 & 52  & This paper \\
Abell 2218 a     & 5.576 & 6.2\tablenotemark{d}
              & $\sim$70\tablenotemark{e}  & 1 \\
LAE J1044$-$0130 & 5.687 & 1.5 & 340 & 2 \\
J123649.2+621539 & 5.190 & 3.0 & 280 & 3 \\
\enddata
\tablenotetext{a}{Observed Ly$\alpha$ flux in units of
$10^{-17}$ ergs s$^{-1}$ cm$^{-2}$.
}
\tablenotetext{b}{Corrected FWHM of Ly$\alpha$ flux in units of
km s$^{-1}$.
}
\tablenotetext{c}{1. Ellis et al. (2001), 2. Ajiki et al. (2002),
and 3. Dawson et al. (2002).
}
\tablenotetext{d}{In Ellis et al. (2001), the Ly$\alpha$ fluxes obtained
with both Keck LRIS and ESI are given. The Ly$\alpha$ flux
given in this table is their average.
}
\tablenotetext{e}{Observed line width, 2.24 \AA, is roughly 
estimated from Fig. 3 in Ellis et al. (2002). After correcting the instrumental
lien width, 1.25 \AA, we obtain the corrected FWHM $\sim 70$ km s$^{-1}$.
}
\end{deluxetable}

%\vspace{0.5cm}

We would like to thank T. Hayashino and 
the staff at both the Subaru and Keck Telescopes
for their invaluable help.
We would like to thank Paul Shapiro and Renyue Cen for their
encouraging discussion on high-$z$ Ly$\alpha$ emitters.
We would also like to thank Zoltan Haiman and an anonymous referee
for their useful comments. 

%-------------------------------------------------------------------------

%------------------------------------------------------------------
%\vspace{1cm}

\begin{figure}
\epsfysize=13cm \epsfbox{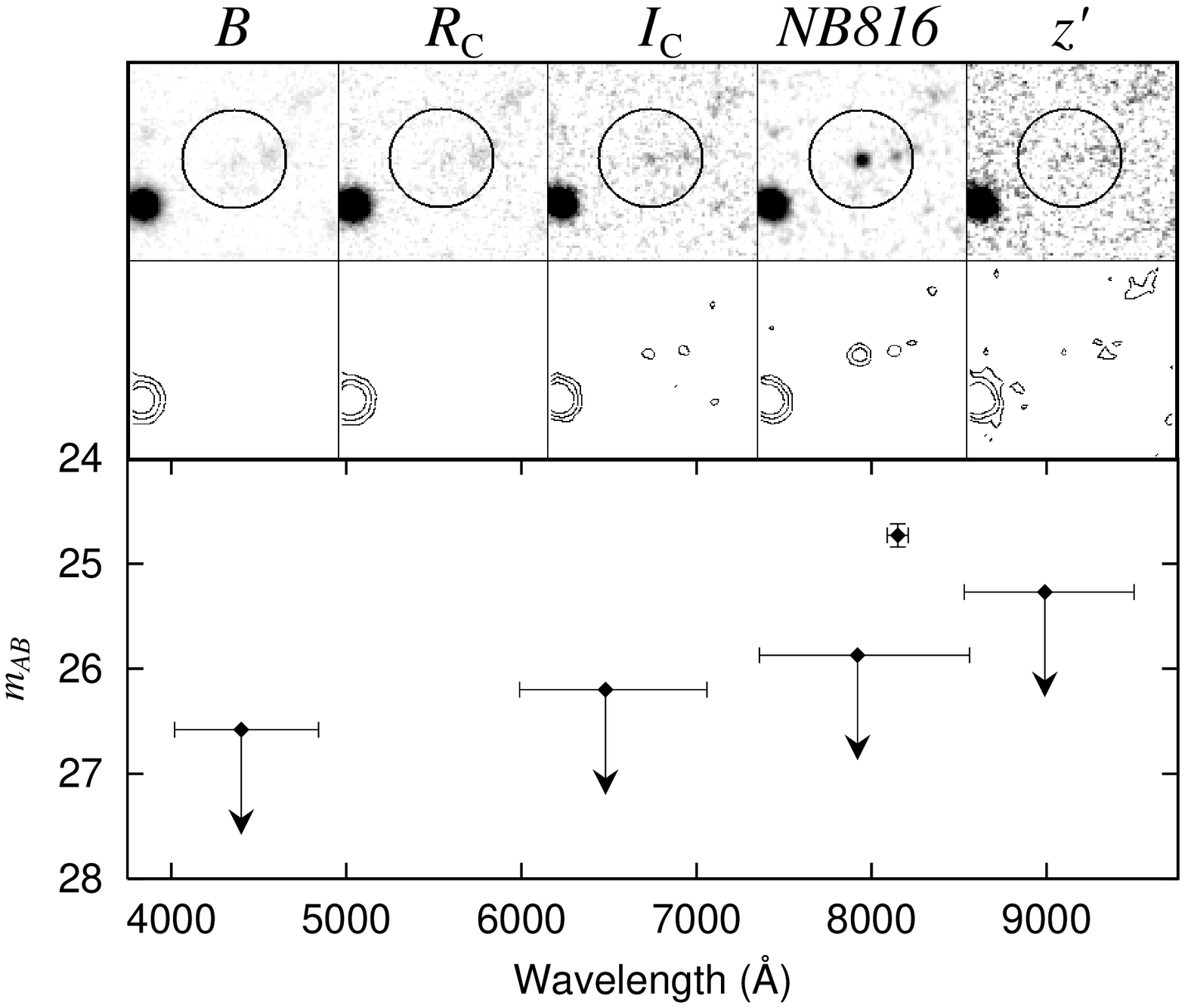}
\caption[]{
Thumb-nail images of LAE J1044$-$0123 (upper panel).
The angular size of the circle in each panel corresponds to 8 arcsec.
Their contours are shown in the middle panel. The lower panel shows
the spectral energy distribution on a magnitude scale.
\label{fig1}
}
\end{figure}

\begin{figure}
\epsfysize=16cm \epsfbox{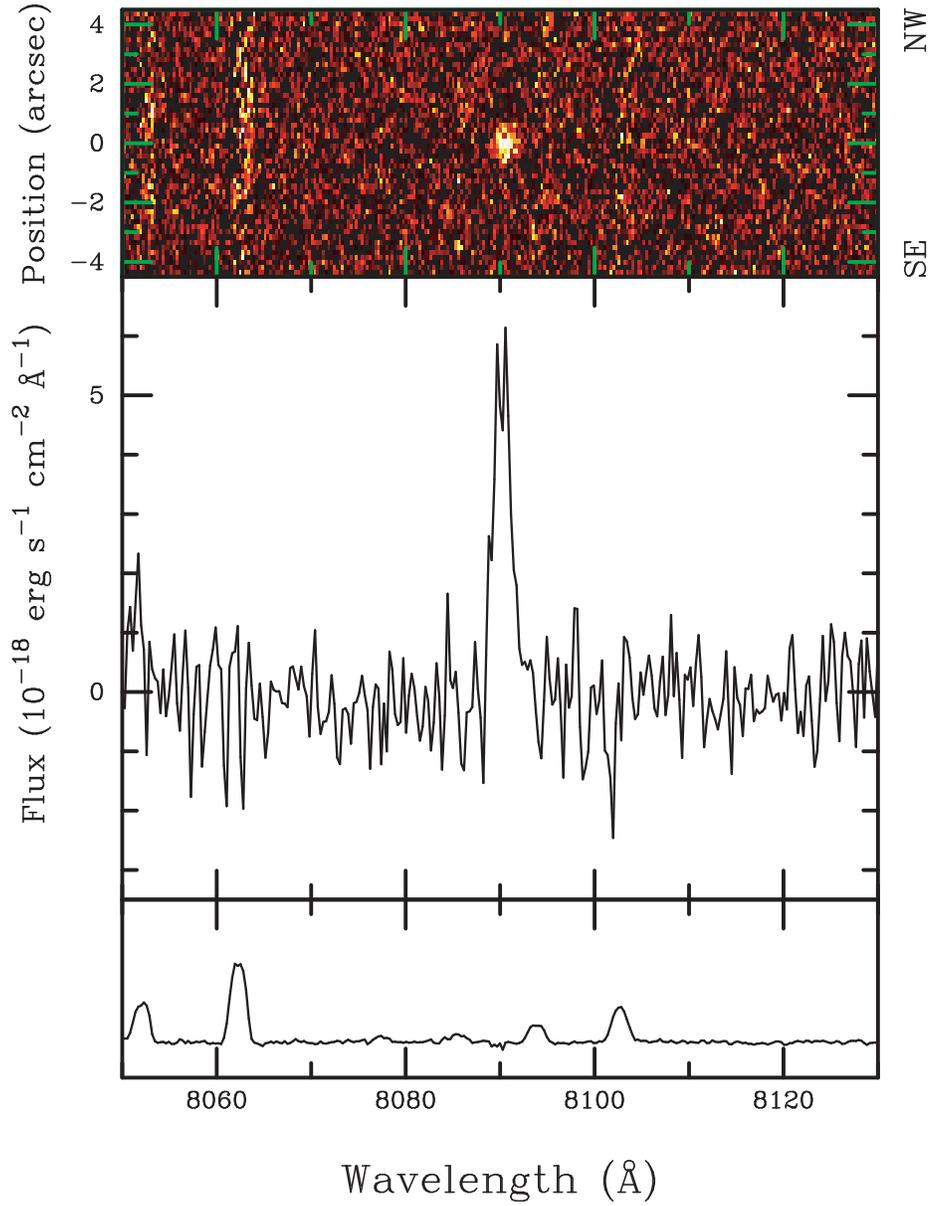}
\caption[]{
The optical spectrum of LAE J1044-0123 obtained
with the Keck II Echelle Spectrograph and Imager (ESI).  We show the
spectrogram in the upper panel and the one-dimensional spectrum
extracted with an aperture of 1.2 arcsec.
\label{fig2}
}
\end{figure}

\end{document}